\documentstyle[12pt,aasms4,epsfig]{article}

\lefthead{Ferreira \& Magueijo}
\righthead{Cosmic Microwave Background}

\begin{document}
\tighten
\title{Evidence for non-Gaussianity in the DMR Four Year Sky Maps} 
\author{Pedro G. Ferreira $^1$, Jo\~{a}o Magueijo$^2$, and  Krzysztof M. 
G\'orski$^{3,4}$}
\affil{$^1$Center for Particle Astrophysics, University of California, 
Berkeley, CA94720, USA\\$^2$Theoretical Physics, 
Imperial College, Prince Consort Road, London SW7 2BZ, UK\\$^3$Theoretical 
Astrophysics Center, Juliane Maries Vej 30,
DK-2100, Copenhagen \O, Denmark\\
$^4$ Warsaw University Observatory, Warsaw, Poland}

\begin{abstract}
We introduce and study the distribution of an estimator for
the normalized bispectrum of the Cosmic Microwave Background (CMB)
anisotropy. 
We use it
to construct a goodness of fit statistic to test the 
coadded 53 and 90 GHz {\it COBE}-DMR 4 year maps for
non-Gaussianity. Our results indicate that 
Gaussianity is ruled out at the confidence level in excess of  98$\%$.
This value is a lower bound, given all the investigated systematics.
The dominant non-Gaussian contribution  is found near the multipole
of order $\ell=16$. Our attempts to explain this effect 
as caused by the diffuse foreground emission from the Galaxy have failed.
We conclude that unless there  exists a microwave foreground emission
which spatially correlates  neither with the DIRBE nor Haslam maps,  
the cosmological CMB anisotropy is genuinely non-Gaussian.
\end{abstract}


\section{Introduction}
We shall consider fluctuations in the CMB as a random field on the
sphere, $\frac{\Delta T}{T}({\bf n})$. One can expand such a field
in terms of Spherical Harmonic functions:
\begin{eqnarray}
\frac{\Delta T}{T}({\bf n})=\sum_{\ell m}a_{\ell m}Y_{\ell m}({\bf n})
\label{almdef}
\end{eqnarray}
For a statistically isotropic field one has
\begin{eqnarray}
\langle a_{\ell_1 m_1}a^*_{\ell_2 m_2}\rangle=C_{\ell_1} 
\delta_{\ell_1 \ell_2}
\delta_{m_1 m_2}
\label{defiso}
\end{eqnarray}
We can also define the two-point function in terms 
of $\frac{\Delta T}{T}({\bf n})$.
Isotropy implies that the correlation matrix can only depend on the
angle between the two points considered. This is encoded in the 2-point
correlation $C^{(2)}(\theta)$. From (\ref{almdef}) and 
(\ref{defiso}) we find
\begin{equation}\label{c2cl}
  C^{(2)}(\theta)={\sum _\ell}{2\ell+1\over 4\pi}C_\ell 
P_\ell(\cos\theta)
\end{equation}
Hence the $C_\ell$ may be regarded as a Legendre transform of the
2-point correlation function. 

It is a standard lore that, barring some mathematical obstructions, one can
reconstruct the probability distribution function of any random field 
from its 
moments. Isotropy imposes ``selection rules'' on these moments. 
For instance, the 3-point moment is given by
\begin{eqnarray}
\langle a_{\ell_1 m_1}a_{\ell_2 m_2} a_{\ell_3 m_3}\rangle=
\left ( \begin{array}{ccc} \ell_1 & \ell_2 & \ell_3 \\ m_1 & m_2 & m_3
\end{array} \right ) C_{\ell_1\ell_2\ell_3}
\label{defnpoint}
\end{eqnarray}
where the $(\ldots)$ is the Wigner $3J$ symbol. The coefficients
$C_{\ell_1\ell_2\ell_3}$ are usually called the bispectrum.
If we assume 
that there are no correlations between different $\ell$ multipoles then
the only non-zero component of the bispectrum is
$C_{\ell\ell\ell}=B_\ell$. The
collapsed 3-point correlation function $C^{(3)}(\theta)$ 
(the average of a temperature squared at one point, and a temperature 
at another point, separated by  an
angle $\theta$) is now
\begin{eqnarray}
C^{(3)}(\theta)={\sum _\ell}{\left(2\ell+1\over 4\pi\right)}^{3/2}
\left ( \begin{array}{ccc} \ell & \ell & \ell \\ 0 & 0 & 0
\end{array} \right )B_\ell P_\ell(\cos\theta)
\end{eqnarray}
in analogy with (\ref{c2cl}). Hence the $B_\ell$ is related to the Legendre
transform of $C^{(3)}$.
The angular power spectrum $C_\ell$ is 
often considered a more powerful tool than 
the correlation function $C^{(2)}(\theta)$ for discriminating
between theories, and one might argue the same way with  regard to the
reduced bispectrum $B_{\ell}$ and the 3-point function $C^3(\theta)$.

The importance of higher order statistics for characterizing
large scale structure has been stressed before (\cite{lss}).
The non-linear evolution of primordial Gaussian fluctuations
has been analysed in  detail (\cite{lss,bouch92}) and 
the skewness arising
in such models has been shown to be consistent with current
observations (\cite{bouch93,gaz94}). 
\cite{luo94} discussed the statistical properties and detectability
of the bispectrum for a variety of non-Gaussian signals.
\cite{kog96a} measured the pseudocollapsed and equilateral
three point function of the DMR four year data and found them to
be consistent with Gaussianity. The analysis performed here
should be considered complementary to that of \cite{kog96a}: non-Gaussian
signals which may be obscured in real space can become evident
in $\ell$ space.

In this letter we shall use a  general 
formalism for generating estimators of higher order moments on a sphere
(\cite{fergormag}). In
this formalism one considers all possible tensor products 
of $\Delta T_\ell$ (each multipole component of the field) 
and from these one extracts the singlet (invariant) term.
In the case of  bispectrum
one has
\begin{eqnarray}
{\hat B}_\ell&=&\alpha_\ell\sum_{m_1m_2m_3}\left 
( \begin{array}{ccc} \ell & \ell & \ell \\ m_1 & m_2 & m_3
\end{array} \right ) a_{\ell m_1}a_{\ell m_2} a_{\ell m_3}
\nonumber \\
\alpha_\ell&=&\frac{1}{(2\ell+1)^{\frac{3}{2}}}\left (
\begin{array}{ccc} \ell
 & \ell & \ell \\ 0 & 0 & 0
\end{array} \right )^{-1}
\label{bispec}
\end{eqnarray}
Note that only even values of $\ell$ lead to nonzero values
of the ${\hat B}_\ell$ due to the symmetries of the Wigner 3-J coefficients.
In practice it is essential to factor out the power spectrum from
our statistic.
We also wish to define statistics which
are invariant under parity transformations, and not just
rotations. Therefore we define $I^3_\ell$ to be
\begin{eqnarray}
I^3_\ell &=&\left| { {\hat B}_\ell\over ({\hat C}_\ell)^{3/2}}
\right| \label{defI}
\end{eqnarray}
where ${\hat C}_\ell=\frac{1}{2\ell+1}\sum_m|a_{\ell m}|^2$.
Our statistics are dimensionless and are normalized so that
a cylindrically symmetric multipole has $I^3_\ell=1$.

The $\ell=2$ case was discussed  and given a physical
interpretation in \cite{mag1}. The quadrupole has 5
degrees of freedom. Of these only 2 are rotationally invariant.
One is the quadrupole intensity $C_2$, and tells us how much
power there is in the quadrupole. The other is essentially
$I^3_2$ and tells us how this power is distributed among the
different $a_{2 m}$ but only as far as there is a rotationally
invariant meaning to the concept. For instance if $I^3_2=1$
then there is a frame in which all the power is concentrated
in the $m=0$ mode. Such a quadrupole is cylindrically
symmetric, but of course the symmetry axis orientation
is uniformly distributed, to comply with statistical
isotropy. If $I^3_2=0$ then on the contrary cylindrical
symmetry is maximally broken. 
The probability distribution
function of $I^3_2$  is uniform in Gaussian theories
(\cite{mag1}).

\section{Goodness of fit and evidence for non-Gaussianity}
\label{chi2}
We will be testing the  inverse noise variance weighted, average maps of 
the 53A, 53B, 90A and 90B {\it COBE}-DMR channels, with monopole
and dipole removed, at resolution 6, in ecliptic 
pixelization. We use the  
extended galactic cut of \cite{banday97}, and 
\cite{benn96} to remove most of the emission from the plane of the Galaxy.
We apply our statistics to the DMR maps before and after correction
for the plausible diffuse foreground emission outside the galactic plane
as described in
\cite{kog96b}, and \cite{COBE}. 
To  estimate the $I^3_\ell$s we set
the value of the pixels within the galactic cut to 0 and 
the average temperature {\it of the cut map} to zero. 
We then integrate the
map multiplied with spherical harmonics  to obtain the estimates of the
$a_{\ell m}$s and apply equations \ref{bispec} and \ref{defI}.

\begin{figure}
\centerline{\psfig{file=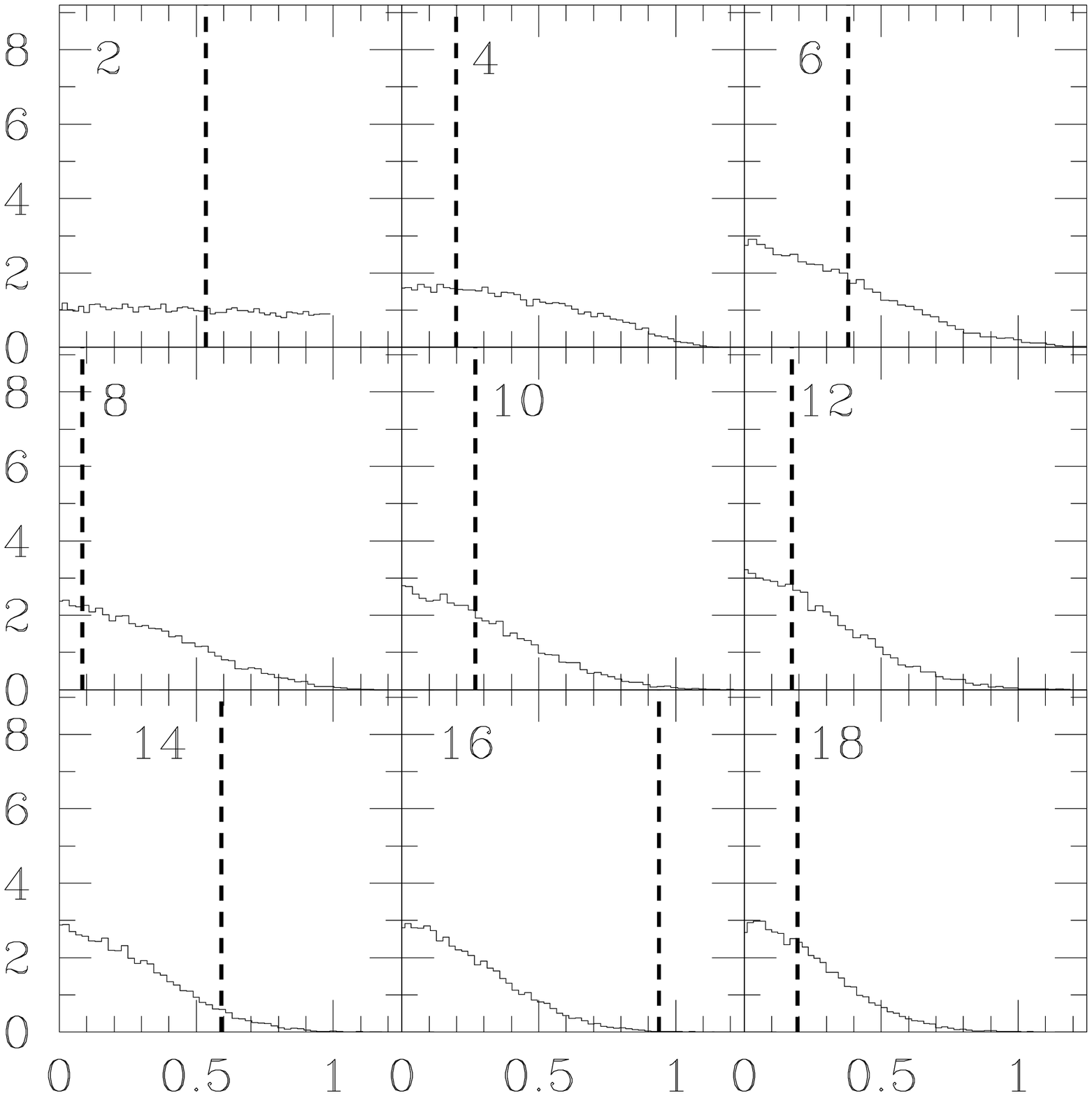,width=12cm}}
\caption{The vertical thick dashed line represents the value 
of the observed
$I^3_\ell$.  The solid line is the probability distribution function
of $I^3_\ell$ for a Gaussian sky with extended galactic cut and
DMR noise.}
\label{fig1}
\end{figure}

We have used Monte Carlo simulations to find the distribution 
of the estimators $I^3_\ell$ as applied to Gaussian maps subject to
DMR noise and galactic cut (see Fig.~\ref{fig1}). These distributions
are very non-Gaussian.  In principle this would complete 
the theoretical work required for converting the observed $I^3_\ell$ 
(which we also plot in Fig.~\ref{fig1}) into a statistical 
statement on Gaussianity, but we proceed further by
defining  a new ``goodness of fit'' statistic as follows.

We wish to construct a tool similar to the $\chi^2$ (often 
used for comparing predicted and observed $C_\ell$ spectra) 
but adapted to the  non-Gaussian distributions $P(I^3_\ell)$.
First, however, recall that if the $\{I^3_\ell\}$ were a set
of $N$ independent, $N(\mu_\ell,\sigma_\ell)$-distributed 
variables, the usual definition of the chi squared would read
\begin{equation}
X^2={1\over N}{\sum{(I^3_\ell-\mu_\ell)^2\over \sigma_\ell^2}},
\end{equation}
with $X^2$ distributed as $\chi^2_N$,
a good fit  represented by $X^2\approx 1$, and
$X^2\ll 1$ ($X^2\gg 1$) corresponding to the 
unusually large (small) scatter in the data given the assumed
variances.
The distribution of $X^2$
is used to find the probability, given the model,
of a value of $X^2$ as large or as small as the one observed.
The converse probability is the confidence level 
for rejecting  the model.

Since the $I^3_\ell$ distributions are  non Gaussian  we 
generalize the $\chi^2$ for a set of probability functions
$P_\ell(I^3_\ell)$ associated with observations $\{I^3_\ell\}$ 
by defining the following functional
\begin{equation}\label{presc}
X^2={1\over N}{\sum_\ell X_\ell^2}=
{1\over N}{\sum_\ell (-2\log P_\ell(I^3_\ell) 
+ \beta_\ell),}
\end{equation}
where the constants $\beta_\ell$ are defined so that for each term
of the sum $\langle X_\ell^2\rangle=1$. The definition reduces
to the usual $X^2$ for Gaussian $P_\ell$. 

As an illustration let 
us first approximate the distributions of the $I_\ell^3$
by $P(I^3_\ell)=2(1-I^3_\ell)$ --- 
a good approximation for $\ell$ around 10.
Then $X^2=-2\log(1-I^3_\ell)$. 
Like the standard $X^2$ one has $0<X^2\ll 1$ for observations close
to the peak of the distribution, here at $I^3_\ell=0$. Indeed
$X^2(0)=0$. However the peak of $P(I^3_\ell)$ is far from its average, 
and so the standard $X^2$ would produce $X^2=0$ at the wrong observation.
For observations far 
from the peak of the distribution (but subject to the constraint
$I^3_\ell\le 1$) $X^2$ goes to infinity. In contrast the standard
$X^2$ would always remain finite.

The proposed $X^2$ therefore does for these non-Gaussian distributions 
what the 
usual $X^2$ does for normal distributions. To illustrate its efficiency
let us find its distribution. First note that
$P(X^2)=\exp(-X^2)$. Consider now a 
$X^2_N$ built from averaging the $X^2$ of $N$ independent
observations:
\begin{equation}
X^2_N={-2\over N}{\sum_\ell \log(1-I^3_\ell)}
\end{equation}
Using characteristics its distribution may be found to be
\begin{equation}
F(X^2_N)={N^N\over (N-1)!}e^{-NX_N^2}X_N^{2N-2}
\end{equation}
This is a $\chi^2_{2N}$. Even if the original invariants were
Gaussian distributed the standard $X^2$ would only be a $\chi^2_N$.
Of course for values of $\ell$ away from 10 the $P(I^3_\ell)$ are
very different from the analytical fit 
presented. In particular for $\ell=2$ the distribution is uniform,
meaning that any observation is compatible with Gaussianity.
Applying the prescription (\ref{presc}) we have that indeed 
$X^2(I^3_2)=1$.

In practice we build a $X^2$ for the {\it COBE}-DMR data by means of Monte
Carlo simulations. We proceed as follows. First we compute the 
distributions $P(I^3_\ell)$, for $\ell=2,\dots,18$, 
for a Gaussian process as measured subject to our galactic
cut, and pixel noises. These $P(I^3_\ell)$ were inferred 
from 25000 realizations (see Fig.~\ref{fig1}). 
From these distributions we then build 
the $X^2$ defined in ($\ref{presc}$), 
taking special care with the numerical
evaluation of the constants $\beta_\ell$. We call 
this function $X^2_{COBE}$.
We then find its distribution $F(X^2_{COBE})$
from 10000 random realizations.  This is very well approximated by 
a $\chi^2$ distribution with 12 degrees of freedom. If all $P(I^3_\ell)$
were as in the analytical fit above, we could conclude that we
successfully measured an effective number of useful invariants
equal to 6. This is less than the number of invariants we actually
measured (10) and this is simply due to 
anisotropic noise and galactic cut. However, had
 we used a standard $\chi^2$ statistics
the effective number of useful invariants would be only 3.

We then compute $X^2_{COBE}$ with the actual observations and find
$X^2_{COBE}=1.81$. One can compute $P(X^2_{COBE}<1.81)= 0.98$.
Hence, it would appear that we can
reject Gaussianity at the $98\%$ confidence level.

\section{Discussion}
The result that we have obtained raises a number of questions
which we shall attempt to answer. From Fig.~\ref{fig1}
it is clear that $I^3_{16}$ is far in the tail of
the Gaussian ensemble and it dominates the statistic.  One would like
to understand the importance of both cosmic variance and noise
to this measurement. We would also like to assess the extent to
which a galactic foreground contaminant could be responsible for this result.

In order to answer the first question we look for Bayesian estimates
for the $I^3_\ell$ as they are {\it for our sky}. To do this we first 
estimate what the temperature fluctuations 
$T_i=\frac{\Delta T}{T}({\bf n}_i)$ in each pixel $i$ in our dataset
are likely to be, given DMR observations $O_i$, and noises
$\sigma_i^2$. We construct the posterior $P(T_i|O_i)$ assuming 
uniform priors in $T_i$, and also that
a priori no correlations exist between the $T_i$. The latter assumption
is often used in image restoration algorithms, such as 
maximum entropy methods. We then produce an ensemble of skies with the
distribution $P(T_i|O_i)$. From it we infer $P(I^3_\ell|O_i)$,
the distributions for what the
$I^3_\ell$ for our sky are likely to be given DMR observations and
noise. This procedure will allow us to assess the importance of noise
in each of our measurements. However note that this analysis is totally
decoupled from the result in the previous section where all we need
to know are the observed $I^3_\ell$, not their estimates for our sky.

\begin{figure}
\centerline{\psfig{file=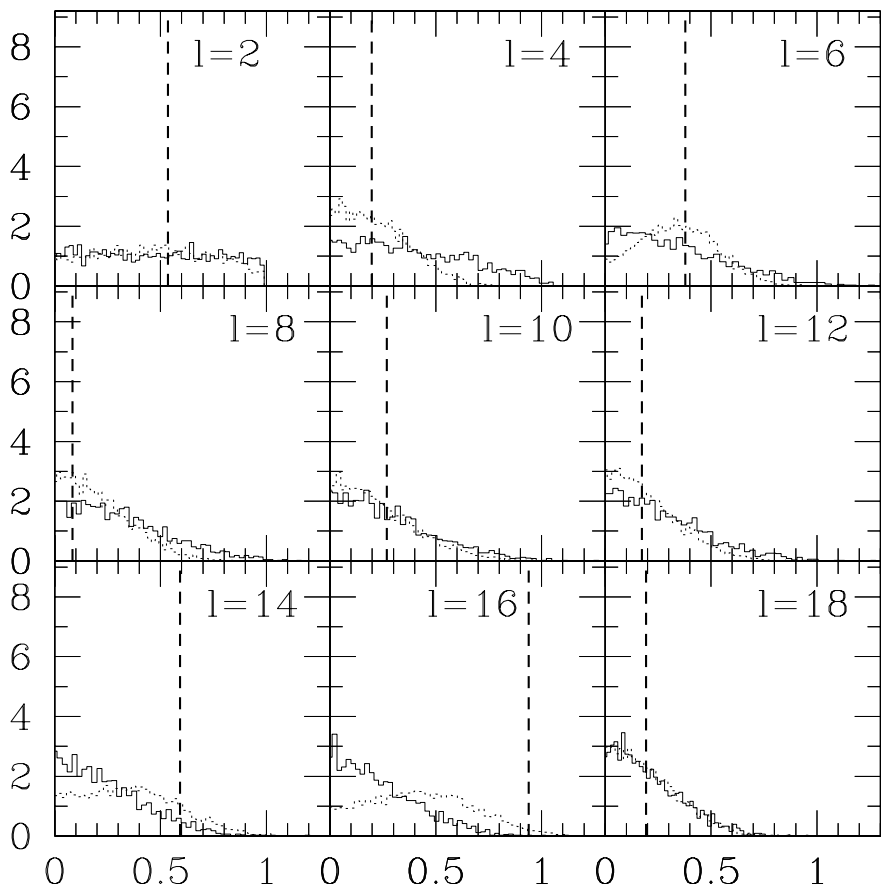,width=12cm}}
\caption{
The vertical thick dashed line represents the value of the observed
$I^3_\ell$, that is $I^3_\ell(O_i)$. The dotted line
is the distribution of where the $I^3_\ell$ {\it for our sky} 
are likely to be, given the observations $O_i$ and  noise,
that is $P(I^3_\ell|O_i)$. The solid line is the cosmic 
variance distribution 
of $I^3_\ell$ for a Gaussian process (subject to the same cut, but 
with no noise).}
\label {fig2}
\end{figure}

In Fig.~\ref{fig2} we plot in dotted lines $P(I^3_\ell|O_i)$
for our data set. We also plot in solid lines
the cosmic variance distribution of $I^3_\ell$ in skies with the 
same galactic cut. The vertical line is the observed invariant
$I^3_\ell(O_i)$.
As expected we see that, as $\ell$ gets larger, the spread in 
$P(I^3_\ell|O_i)$ due to noise becomes more important,  at $\ell=18$ 
dominating the distribution function. 
On the other hand we clearly have succeeded in making measurements for 
$\ell=4,6,8,12,14,16$. For them $P(I^3_\ell|O_i)$
are peaked and clearly different from the cosmic variance distribution. 
The fact that $P(I^3_{16}|O_i)$ does not peak at $I^3_{16}(O_i)$
is merely a failure of the prior. The measurement of
$I^3_{16}(O_i)$ is therefore a signal and is not dominated by noise.
We have further checked that the signal to noise in power 
at $\ell=16$ is of order 1.

Next we wish to know if galactic emissions could be blamed for 
this result. We can proceed in three ways. Firstly we may use instead
the DMR cosmic emission maps, where a linear combination of the 
various DMR channels is used to separate out the foreground 
Galactic contamination. In these maps the noise level is considerably 
higher. Plotting the counterpart of Fig.~\ref{fig2} for this case
we find that
the distributions of 
the actual $I^3_\ell$ for our sky, given noise induced errors, are
very similar to their cosmic variance distributions. 
The measurement is therefore dominated by noise and inconclusive.
We find $X^2_{COBE}=.4$, consistent with Gaussianity, 
but this is a mere check of the Gaussianity of noise. 
Hence this approach towards foregrounds turns into a dead end, but serves
to show how large angle Gaussian tests is a field constrained 
by noise, not cosmic variance.

\begin{figure}
\centerline{\psfig{file=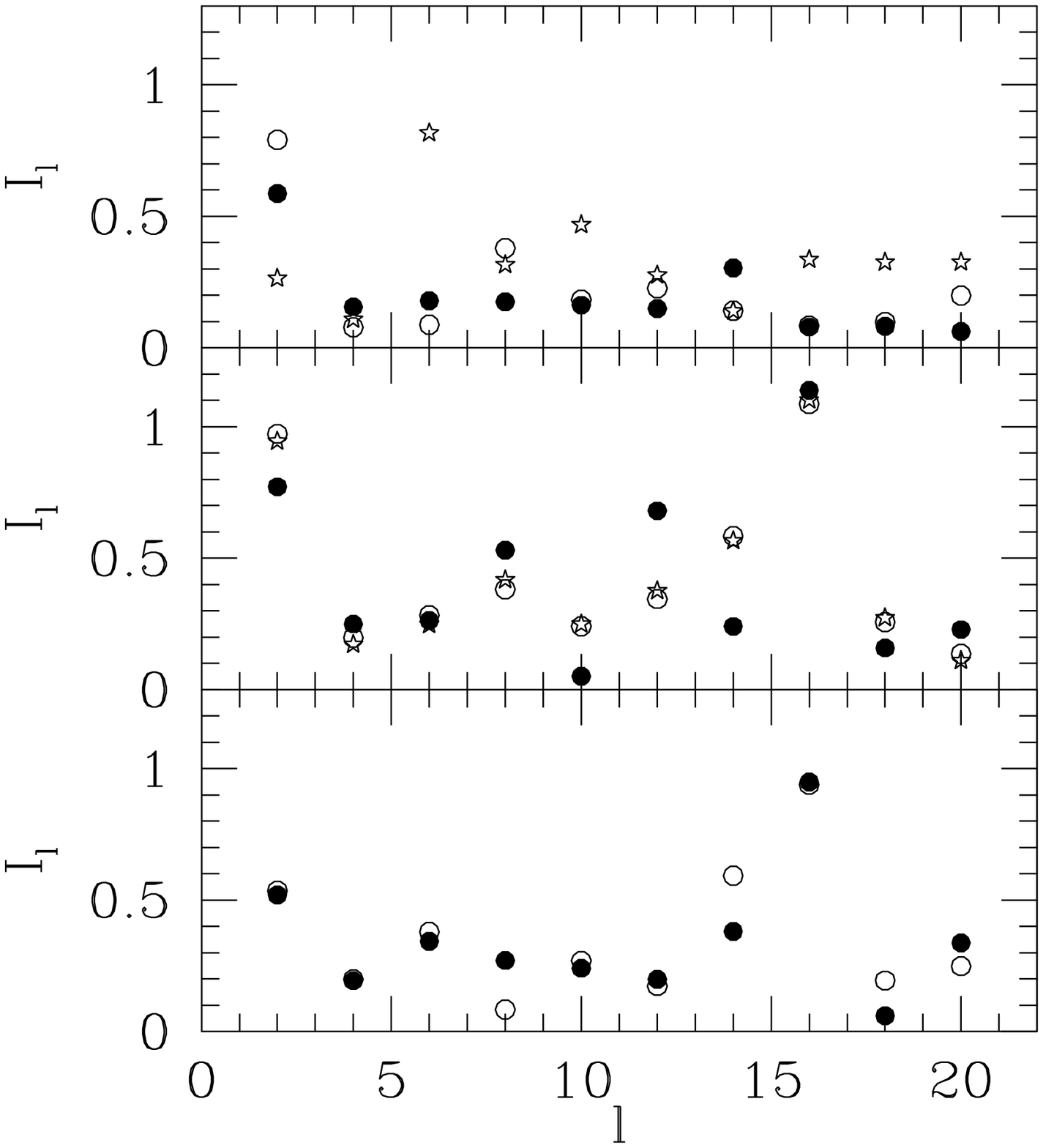,width=12cm}}
\caption{In the bottom panel we plot the measured $I_\ell$ from the
 data in ecliptic coordinates (open circles) and
galactic coordinates (solid circles);  the center panel has
the foreground corrected data in ecliptic coordinates (open circles)
and galactic coordinates (solid circles) and ecliptic coordinates with
galactic frame coupling (stars); the top panel has the
DIRBE 100$\mu$m (open circles ) DIRBE 240$\mu$m (solid circles ) 
and Haslam (stars ) data. }
\label{fig3}
\end{figure}

As an alternative approach 
we may subject galactic templates to the same analysis.
At the observing frequencies the obvious contaminant should be foreground
dust emission. The DIRBE  maps (\cite{boggess92}) supply us with a
useful template on which we can measure the $I^3_\ell$s. We have done
this for two of the lowest frequency maps, the $100$ $\mu$m  and the
$240$ $\mu$m maps. The estimate is performed in exactly the
same way as for the DMR data (i.e. using the extended Galaxy cut).
We performed a similar exercise with the Haslam 408Mhz (\cite{haslam}) map. 
We display their values in Fig.~\ref{fig3}.
As expected the two maps have consistent values
for the $I^3_\ell$. However they do not have a non-Gaussian value at
$\ell=16$. Indeed for all $\ell$ the $I^3_\ell$ are within 
Gaussian cosmic
variance error bars. This is not surprising. DIRBE maps exhibit structures
on very small scales. These should average into a Gaussian field when
subject to a $7^\circ$ beam.

As a third alternative we may use foreground corrected maps.  In these one
corrects the coadded 53 and 90 Ghz maps for the DIRBE correlated emission.
We have considered corrected maps in ecliptic and galactic 
frames, and also another map made in the ecliptic frame but with the DIRBE
correction forced to have the same coupling as determined in the galactic
frame. As shown in Fig.~\ref{fig3}, 
in all of these the non-Gaussian signal at $\ell=16$ is enhanced,
although we observe large variations
in $I^3_\ell$ at $\ell=4-8$ (a phenomenon noticed before  when estimating 
$C_\ell$-s). In fact the corrected maps exclude Gaussianity at the
confidence level of 99.5\%.

It would be interesting to relate our result to 
the curious dip in power at $\ell\approx 16$ provided by the maximum
likelihood estimates in \cite{gorski97}. These show that, 
{\it assuming a Gaussian signal}, the power in signal and noise is unusually 
low at $\ell\approx 16$. One wonders how this would be affected if non-Gaussian
degrees of freedom were allowed into the estimation (\cite{fergormag}).

We have also subjected our work to a variety of numerical tests.
Arbitrary rotations of the coordinate system affects results
to less than a part in $10^5$. More importantly, comparing data pixelized
in the ecliptic and galactic frames, we found that our results were very
robust, indeed  more so than the power spectrum estimation (see
the bottom pannel of Fig.~\ref{fig3}).
We also tried different
galactic cuts, and found that although the non-Gaussian signal 
gets transferred
to other $\ell$, one does not fully erase it until a cut of $\pm 40^\circ$ 
is applied. Finally we checked the effect of varying the offset
in the cut map. We found that for any other prescription than the 
one used the effect is enhanced, often leading to rejecting Gaussianity 
at more than the 99.5\% confidence level.

To conclude, we have not been able to attribute our result to 
a known contaminating source or a systematic. Indeed the confidence
level quoted refers to the worst result obtained within the
set of effects explored. 
Of course it is always possible that this non-Gaussian 
signal comes from some yet unmapped foreground, which cannot be 
separated from the CMB anisotropy signal in the {\it COBE}-DMR data ---
the poorly known free-free emission from the Galaxy comes to mind here.

If indeed our results  are due to a foreground contamination one should note 
the following two points.
First, we would have demonstrated that DMR data is
more contaminated by foregrounds than thought before. Second,
Galactic emissions on the scales considered are often
assumed to be Gaussian. In fact this assumption is used in  
subtraction algorithms based on the idea of optimal filtering. 
The discovery of a distinctly non-Gaussian galactic emission
would in the very least require a rethinking
of the foreground subtraction algorithms. 

If, on the other hand, the CMB signal 
itself is demonstrably non-Gaussian, we would not  need to over-emphasise the
epistemological implications of our findings.

\section*{Acknowledgements}
We thank T. Banday, T. Bunn,  
K. Baskerville, 
M. Hobson, A.Jaffe, J.Nunes, D.Spergel  and R.Taillet.
Resources of Starlink (IC) and of the COMBAT collaboration 
were used in this project.
PGF was supported by NSF (USA), JNICT (PORTUGAL) and CNRS (FRANCE),
JM by the Royal Society and TAC, and KMG by Danmarks
Grundforskningsfond (TAC) and partly by NASA-ADP grant.

\newpage

\begin{thebibliography}{}

\bibitem[Banday et al 1997]{banday97} Banday, A.J.,
G\'orski, K.M., Bennett, C.L., Hinshaw, G., Kogut, A.,
Lineweaver, C.,  Smoot, G.F., and Tenorio,  L. (1997) {\it Ap.J.}, 
{\bf 475}, 393

\bibitem[Bennet et al 1996]{benn96} Bennet, C.L., Banday, A.J.,
G\'orski, K.M., Hinshaw, G., Jackson, P.D., Keegstra, P., Kogut, A.,
Smoot, G.F., Wilkinosn, D.T., Wright, E. L. (1996) {\it Ap. J.}
{\bf 464}, 1

\bibitem[Boggess et al. 1992]{boggess92}Bogess et al. (1992) {\it Ap. J}
{\bf 397} 420

\bibitem[Bouchet et al. 1992]{bouch92} Bouchet, F.R., Juskiewicz, R.,
Colombi, S., Pellat, R. (1992) {\it Ap. J. Lett.} {\bf 394}, L5

\bibitem[Bouchet et al. 1993]{bouch93} Bouchet, F.R., Strauss, M.,
Davis, M., Fisher, K.B., Yahil, A., Huchra, J.P. (1993)
{\it Ap. J.} {\bf 417}, 36

\bibitem[Ferreira, G\'orski \& Magueijo 1998]{fergormag} Ferreira, P.~G.,
G\'orski, K.~M., Magueijo,~J. (1998) in preparation

\bibitem[Gazta\~{n}aga 1994]{gaz94} Gazta\~{n}aga, E. (1994)
{\it M.N.R.A.S.} {\bf 268}, L1.

\bibitem[G\'orski et al. 1996]{COBE} G\'orski, K.~M., Hinshaw, 
G., Banday, A.~J., 
           Bennett, C.~L., Wright, E.~L., Kogut, A., Smoot,
           G.~F., Lubin, P. (1996) {\it Ap. J. Lett.} {\bf 464}, L11

\bibitem[G\'orski (1997)] {gorski97} G\'orski, K.~M., in Microwave Background 
Anisotropies,
Proceedings of the XVIth Moriond Astrophysics Meeting, March 1996,
Editions Frontieres 1997;  also astro-ph/9701191 

\bibitem[Haslam (1982)] {haslam} Haslam, C.G.T et al (1982)
{\it Astron. Astrophys. Suppl. Ser.}, {\bf 63}, 205.

\bibitem[Kogut et al. (1996)]{kog96a} Kogut, A.,  Banday, A.~J., 
           Bennett, C.~L., G\'orski, K.~M., Hinshaw, G., Smoot,
           G.~F.,  Wright, E.~L. (1996) {\it Ap. J.} {\bf 464} L29

\bibitem[Kogut et al. (1996)]{kog96b} Kogut, A.,  Banday, A.~J., 
           Bennett, C.~L., G\'orski, K.~M., Hinshaw, G., Smoot,
           G.~F.,  Wright, E.~L. (1996) {\it Ap. J.} {\bf 464} L5

\bibitem[Luo 1994]{luo94} Luo X. (1994) {\it Ap. J.} {\bf 427}
L71.


\bibitem[Magueijo (1995)]{mag1}  Magueijo, J. (1995) 
{\it Phys. Lett.} {\bf
B342} 32-39 ERRATUM-ibid {\bf B352} 499.

\bibitem[Peebles 1980]{lss} Peebles, P.J.E. (1980) {\it The Large
Scale Structure of the Universe} (PUP, Princeton)


\end{thebibliography}
\end{document}